\documentclass{article}
\usepackage{graphicx}
\begin{document}


\title{Between the Information Economy and 
Student Recruitment: Present Conjuncture and Future Prospects}

\author{Fionn Murtagh\footnote{Fionn Murtagh is also Professor of 
Computer Science in 
the University of London.  
Department of Computer Science,
Royal Holloway, University of London,
Egham TW20 0EX, England}
\\
Director, Information, Communications and Emergent Technologies\\
Science Foundation Ireland \\
Wilton House, Wilton Place, Dublin 2, Ireland \\
 \\
Email: fmurtagh@acm.org}

\maketitle

\begin{abstract}
In university programs and curricula, in general we react to the
need to meet market needs.  We respond to market stimulus, or 
at least try to do so.  Consider now an inverted view.  Consider
our data and perspectives in university programs as reflecting 
and indeed presaging economic trends.  In this article I pursue 
this line of thinking.  I show how various past events fit very 
well into this new view.
I provide explanation for why some technology trends 
happened as they did,
and why some current developments are important now.  
\end{abstract}

\section{The Downturn in Academic Computer Science Undergraduate 
Student Recruitment}

The student recruitment crisis of Computer Science and 
Engineering (CS and E)  has been seen 
as one where there is over-provision 
of supply relative to demand.   A response has 
been sought in more public 
outreach and in restructuring course provision.  I am 
completely at one with this 
important work.  

In this article I want to look at this 
context of discomfort and indeed 
of crisis from a very different vantage point.  I will 
argue that we can view the swings 
of fortune in CS and E student recruitment as a prism with 
which to view large scale underlying technology and economic trends.  
I will illustrate this argument in various 
ways.  


In an ideal world we could step back and just note that 
student demand has gone 
elsewhere, assuming relatively unchanging demographics.  Maybe we 
would even retool our expertise, by changing research 
discipline for example.  
But there has been very great fluctuation in student demand
and reacting overly hastily to 
the ups and downs of fortune is rarely a good idea.  

In this article I will look 
closer at this fluctuation in student demand for CS and E.  
I will reverse the usual 
view of trying to explain student demand in terms of
deep-lying  economy needs.  
Instead I will present the view that major fluctuations 
in the economy can  --  up to 
a point -- be interpreted and understood by the available 
data on student demand.   The fit with a wide range of 
important technology trends is very good, as I will exemplify.

Between technological upswings I will present the view that one should 
prepare well for the next upswing.  
In regard to how we prepare for the future, one 
point to be noted is that our perspective will be a cloudy one
if traditional economic 
categories like manufactured goods and services dominate our 
thinking.  See section \ref{econview} for further discussion here.  

Relatively interchangeably in this article I will use the 
terms CS and E, and ICT or information and communications 
technology.  The latter is preferred when the industrial,
commercial and market aspects are strongly represented.

\section{The 
Information Society and the New Economy Periods of Spectacular Growth} 

There have been two major ICT-led economic booms in recent times.  
In both phases, the communications aspect of computing was hugely 
prominent.  

Figure \ref{fig1}  shows an educational 
reflection of what happened and when.  I use North American data a number
of times in this article because it is of high quality and collected
in a consistent way over many years.  Twice, we find major 
upswings in attractiveness of the science and technology.  
Figure \ref{fig1} relates to incoming student intentions.  Like business
confidence surveys vis-\`a-vis the economy, Figure \ref{fig1} 
expresses the pulling power of the discipline (or the 
generally perceived tight
cluster of disciplines associated with CS and E).  
We see, 
well-mirrored in Figure \ref{fig1}, a 
massive take-off of, and interest in, computerization.  By the late 
1980s, this was in free-fall.  Growth was ratcheted up in the 1990s.  
By early 2001, the economy was slipping fast (see e.g.\ \cite{rita} 
in support of the downturn starting in late 2000).  

I will look at these two massive technology upswings, well 
expressed by the bumps in Figure \ref{fig1}.  In line with what 
they have been often called, I will use the respective terms 
of Information Society and New Economy periods or booms.  
As a synonym here for boom, I will use the term upwelling.  
In ocean processes, upwelling is heat- and gravity-engendered.
Upwelling events have important implications for biomass and 
later parts of the food chain.  The upwelling metaphor is an apt 
one.

\subsection{The Information Society Boom}

Periodizing the earlier Information Society boom may be helped by Figure 
\ref{fig1} and this 
note from \cite{vegso}  
that ``between 1980 and 1986, undergraduate CS 
production nearly quadrupled to more than 42,000 degrees. This 
period was followed 
by a swift decline and leveling off during the 1990s''.  

The first great boom was the personal computer (PC) 
led one, focused on the computerization of society, 
and it also saw a great deal of early activity in networking.
This boom was led by the generalized PC uptake in the early 1980s.  
It put to rest 
the debate on whether computerization of society could lead to 
productivity and 
general growth.  A key text, with influence internationally, was 
the Nora/Minc report \cite{nora}, which inspired 
French telecoms through Minitel, for example (an early chapter 
of \cite{nora} is 
entitled ``From informatics to 
telematics'').   Among the very opening lines are: 
``Increasing computerization of society is at the heart of the crisis''
and the economic, 
political and social crisis is characterized generally by 
``grave, new challenges''
under the overall heading of 
``the French crisis of informatics''.  

Tectonic movements lay in the technology undergrowth, 
underlying science and engineering, and in market forces. 
Just to sketch a few important events of the time,
Intel's first microprocessor was launched in November 1971.  
The Apple II personal computer, introduced in 1977, was in continuous
production until 1993.   It was successful and 
mass-produced.  The IBM PC, or IBM 5150, was launched in August 1981.  
Very soon the IBM PC had massively overtaken other alternative 
platforms \cite{ibmpcshare}.  1984 saw the divestiture of AT\&T's 
operating companies into seven Regional Bell Operating Companies 
(see \cite{pennings} for discussion and historical background).
Mobile telephony was launched in the US in 1984.
What was termed the deregularization or liberalization of the 
telecoms market was initiated in the European Union in 1985 
through directives under the Treaty of Rome.  The massive growth
from the early 1990s of mobile telephony relative to fixed line telephony 
is well charted in \cite{pennings}.  So too are the organisational
changes in the sector, including domestic and international alliances,
and mergers and acquisitions (M\&As) all of which hugely increased.

\subsection{The Telecoms View Preceding the Information Society Boom} 

Against a background of market dominance by IBM, and the use of 
videotext in the 
UK (information delivered to end users by television signal), the 
national telecoms 
provider in France, DGT -- Direction G\'en\'erale 
des T\'el\'ecommunications, obtained a 
superministerial budget in 1975, and in 1978, Simon Nora and 
Alain Minc submitted 
their hugely influential report, \cite{nora},
to President Val\'ery 
Giscard d'Estaing.

The Nora/Minc report forecasted (the following is taken from
\cite{rheingold}):    
``A massive social computerization will take place in the future, 
flowing through 
society like electricity. ...  The debate will focus on 
interconnectability.''
The report concluded 
that the advent of cheap 
computers and powerful global communications media was leading to 
``an uncertain 
society, the place of uncountable decentralized conflicts, a 
computerized society in 
which values will be object of numerous rivalries stemming from 
uncertain causes, 
bringing an infinite amount of lateral communication.'' To continue 
to compete in the 
first rank of nations, Nora and Minc exhorted, France would have to 
mount a full-scale 
national effort in the new field they named telematics 
(merging the terms ``telecommunications'' and 
``informatics''). They did not fail to 
note that ``Telematics, 
unlike electricity, does not carry an inert current, 
but rather 
information, that is to say, power'' and that ``mastering the 
network is therefore an 
essential goal. This requires that its framework be conceived in 
the spirit of a public service.''

Officially launched in 1982, Minitel was a great success.  In 
1998 there were 5.6 million Minitel terminals available for this use of this
secure but closed network \cite{mchoes}.

\subsection{Between the Information Society and New Economy 
Booms: An Example from Financial Data and Information}

In this section I will look at one economic sector and how 
a major initiative was
undertaken and grown before and then during the 1990s
New Economy boom.  I use it as an apt 
example of where and how new initiatives can be seeded
to take advantage of economic doldroms, and perhaps particularly 
advantageously during such downbeat periods.  

Financial services now account for a good part of leading 
economies.  In the US, financial services contribute to GDP 
(gross domestic product) at 
8 percent.  In New York City, in 2007 the finance industry was 
``responsible for nearly one third of all wages earned''
\cite{story}.  In the UK, the financial services sector contributes
6 percent to GDP and employs 4 percent of the national workforce.
(See \cite{expert}).  

Finance is based on the direct and immediate processing of 
data and information. In this sense it is one big application 
of ICT.  

The International Financial Services Center, IFSC, was established 
in Dublin in 1987 
between the two boom periods.  It has been a significant success 
story.  The IFSC 
now has 10,700 employed, growing by 1000 per year.  More than 430 
international 
operations trade in the IFSC, and a further 700 are approved to 
carry on business 
there.  
From a very low base at the beginning of the 1990s, Ireland has become an 
established center for the European investment funds industry, as 
shown in Table \ref{tab1} \cite{expert}.  


\begin{table}
\begin{center}
\begin{tabular}{lr}\hline
Luxembourg & 24.4\% \\
France    & 19.7\% \\
Germany   & 13.4\% \\
UK        & 10.3\% \\
Ireland   & 9.5\% \\
Italy     & 5.1\% \\
Spain     & 3.8\% \\
Other    & 13.8\% \\
Total    & 100\% \\ \hline
\end{tabular}
\end{center}
\caption{Percentage net assets of the European investment fund industry, 2006.
From \cite{expert}.}
\label{tab1}
\end{table}

The December 2007 financial services strategy report \cite{expert} 
on the Irish and international financial 
services sector makes interesting reading too that 
links up with the 
growth in PhDs.  (This is discussed further in 
section \ref{phd}.)  This report 
 provided a rationale 
as to why and where more PhDs are needed in this sector.  
Rather than 
``skilled generalists'' lacking specialized knowledge, this report called 
for ``a greater 
focus on specializing in a number of selected areas which would support the 
development of a distinctive competence which was more aligned 
with a mid to high 
cost base.''

\subsection{The New Economy or Dot-Com Boom}

The second great boom came about through the web, with 
complementary activity in 
telecoms, e-commerce and dot-com venture capital and finance
generally.  Wide and popular 
take-up of the web was consolidated with the 
release of the Mosaic browser in early 1993 by Marc Andreessen, 
a student who  
graduated in 1993 at the University of Illinois at Urbana-Champaign.  
This led to 
absolute dominance over other information sharing systems that were 
current in the 
very early 1990s, such as Gopher and Veronica, WAIS
(wide area information system, based on the Z39.50 protocol), 
archie and others.  

Two markers of the Dot-Com boom are to be seen in Ireland and in Finland.

The Celtic Tiger \cite{oriain}
was a term coined by \cite{gardiner} in August 1994. 
The parallel was with the Asian Tiger economies.  
In some years of the Celtic Tiger period growth, 
measured by real GDP (gross domestic product), was more 
than 10\%.  Statistics and discussion can be seen at \cite{esri}.
By 2008, the ICT sector had grown in Ireland to employ more than 
91,000 people.  The Irish software sector alone accounts for
10\% of Ireland gross domestic product \cite{kevinryan}.    
This spectacular Irish growth took off in 1993 and contributed 
crucially to Ireland's impressive growth up to 2001 \cite{fitzgerald}.  
So the 
Irish Celtic Tiger period began at the same time as the popular 
take-up of the web, and both grew in tandem.  

In Finland, the history of Nokia is revealing also from the point 
of view of timing relative to the New Economy.  Nokia evolved from being an 
industrial conglomerate   
dating from the 1860s.  It was  
established as a wood pulp mill in 1865; moved to rubber and cable companies 
operating in 
alliance with a forestry company from 1922 to 1966/1967; following 
a merger it 
expanded into electronics; and from this, telecoms took off in the 1990s.
The take-off of 
Nokia was started in the 1990s, at roughly the same time as the
Irish Celtic Tiger take-off, and the popular upsurge of the web.  Nokia
point to a management decision that in 1994,   
``formulated the key elements of 
Nokia's strategy: leave old businesses and increasingly focus on 
telecommunications.''

So a range of momentous decisions and events were happening at roughly the 
same time, relating to: networks; mobile telecoms; user interfaces;
the information economy; and information distribution.  On the latter, 
information distribution, the July 1994 plunge of comet Shoemaker-Levy
into the planet Jupiter, lasting a week, was an early example.  Networks 
including the young web came of age at that time, through massive 
worldwide interest.  My role included analyzing 
image data and getting information out by all available means -- web, 
other networks (e.g., CompuServe, a dial-up network later absorbed into
AOL), news and television media.  The context is 
described by \cite{whitehouse}.

The Dot-Com boom collapsed by early 2001, and it may be the case 
that we are now 
pulling out of the downturn.  
A good proxy for whether we are or not is the attractiveness manifested by 
undergraduate student recruitment.  Internationally
this has been in freefall since 2001.  
There have been bad consequences: some departments have retrenched, 
and old debates about the nature of our science and engineering have
again developed an unpleasant rawness.  

``After 
six years of declines, the number of new CS majors in fall 2006 was 
half of what it 
was in fall 2000 (15,958 versus 7,798)'' \cite{vegso1}.  
Nonetheless the prognosis 
stated there is hopeful that a turn-around is now happening.  

The response to the crisis of student recruitment in CS, with knock-on 
budgetary 
(salary and support) effects, has been surprisingly uniform.  It has led 
to attempts to 
refocus undergraduate curricula into new digital media such as 
digital music; 
games technologies; and information security.  

\begin{figure}
\includegraphics[width=12cm]{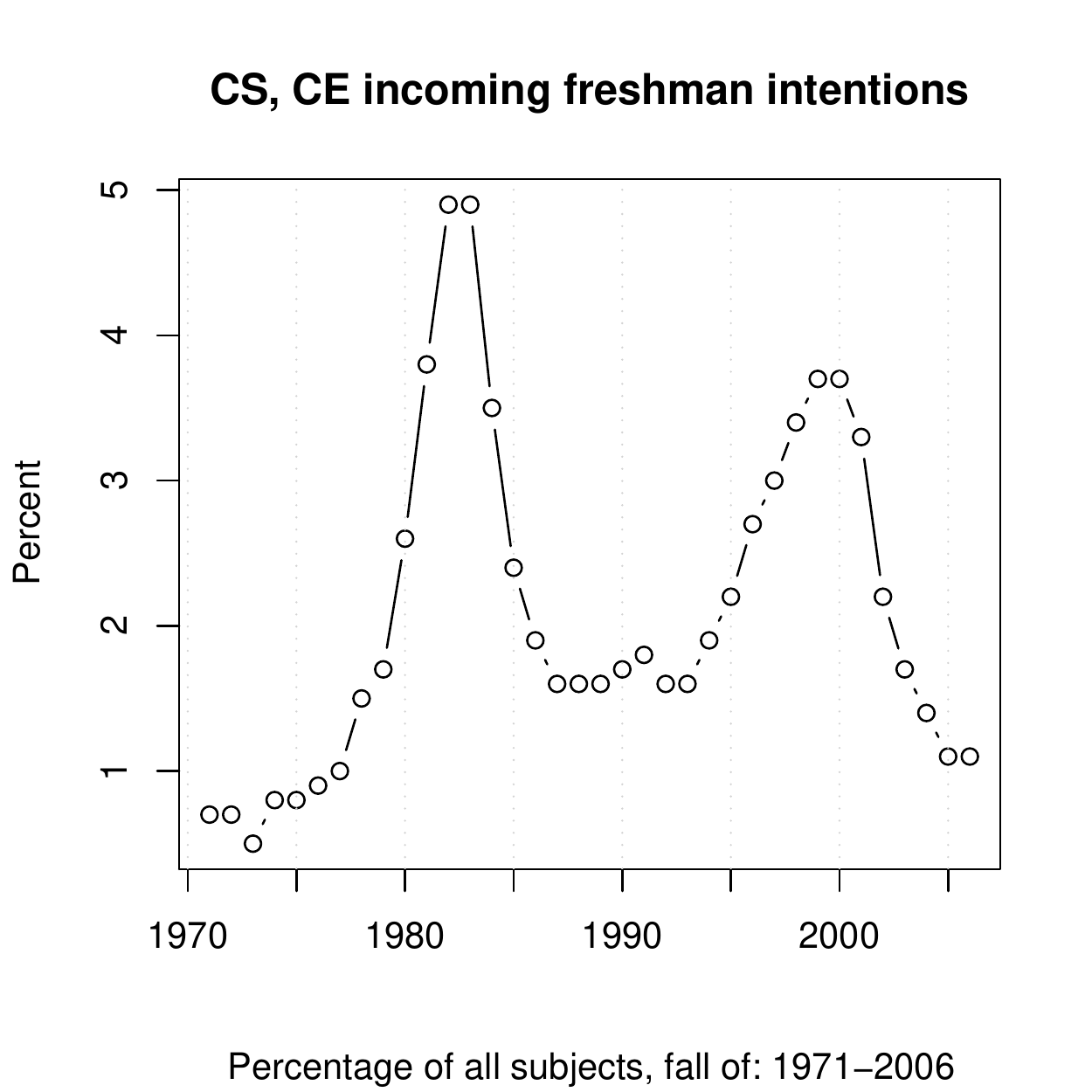}
\caption{A clear view of the two great peaks in 
attractiveness of CS in the past decades.  
These are 
intentions and not actual commitments to degree programs, 
surveyed from incoming freshmen.  Data from 
www.cra.org/wp/index.php?p=104 and also 
www.imageofcomputing/pdf/Heri\_Study\_2006.pdf  Data analyzed by
CRA, Computing Research Association; 
originally from HERI, Higher Education Research Institute,
University of California, Los Angeles. In the discussion I 
associate the two peaks with, respectively, the Information 
Society and the New Economy upwellings.}    
\label{fig1}
\end{figure}
 
\begin{figure}
\includegraphics[width=14cm]{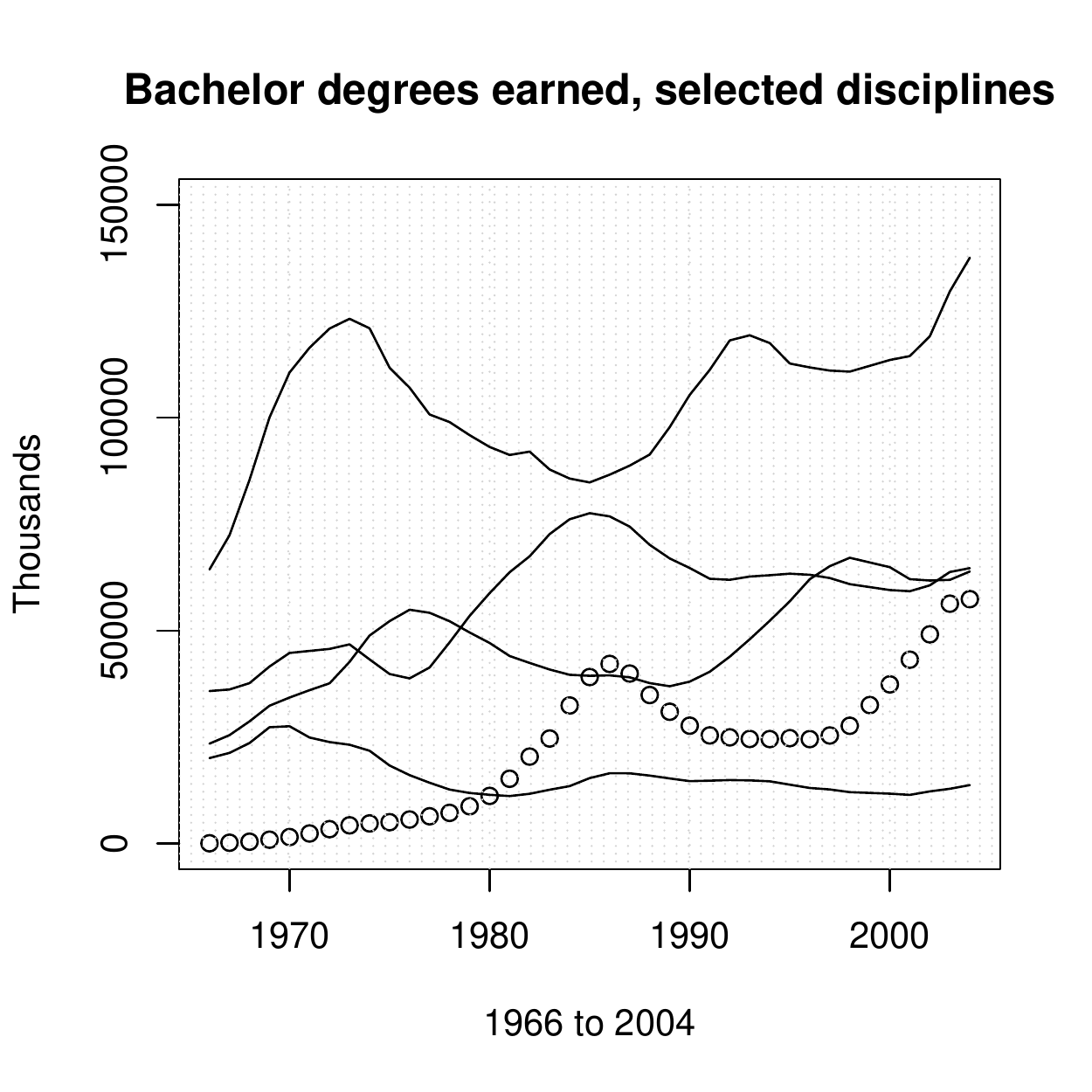}
\caption{The comparative setting.  NSF Division of 
Science Resources data.  See: http://www.nsf.gov/statistics/nsf07307.
Computer Science is the curve that is highlighted through the 
succession of large open dots.  
Other curves from top to bottom: Social Sciences; Engineering
(noticeably sharing the peak of the Information Society boom); 
Biological Sciences; and Mathematics.  Data for the year 1999 was not 
available, so I linearly interpolated.}  
\label{fig3}
\end{figure}
 
One interesting thing about Figure \ref{fig3} 
is that it shows where the students went, 
given their flight from CS and E.   Physical Science (Physics, Chemistry, 
Astronomy, 
Other) had a very similar curve to that of Mathematics, so we do not show 
it here. 
It does appear that for the 1980s boom, CS gained greatly in 
shifting 
students in such a way that ultimately Biological Science and Social 
Sciences were the losers then.   In the 1990s boom there is some
indication of this swing again, albeit less pronounced.  
Note that in Figure \ref{fig3} the 
degrees awarded can be expected to have some lag relative to 
underlying economic developments, and relative to intentions as 
seen in Figure \ref{fig1}.  

\subsection{The Information Society and 
New Economy Upwelling Periods: View from Economics}
\label{econview}

How real the ICT base for society had become remained for long an
open issue to be addressed, -- a set of questions 
rather than a resoundingly clear response
to profound structural changes in economy and society.  It was so
for both the Information Society upwelling of the 1980s and for the  
New Economy of the 1990s.
  
Pakko \cite{pakko} discusses how far there was qualitative change. 
Temple \cite{temple} discusses whether or not the economy had
become structurally new.  
In both cases these authors argue positively.  The very fact that 
such questions were posed is what is curious.

IBM announced its first PC in August, 1981, but the old style of 
economics showed no productivity gains from the PC boom of the 1980s.  
PC sales peaked in 1995.  For PCs, accessories and components, 
``... demand has slowed sharply. If we 
look at nominal order growth for this same industry we see a virtual 
collapse in orders 
between 1994 and 1997 despite some firming in the overall economic 
growth rate.''  \cite{veneroso}.    
The long heralded Information 
Society was criticized as being nowhere to be found \cite{mandel}.
This is all very curious when looked at now in our rear view mirror.  
The low point for long-term productivity growth had later
to be revised to 1982, as 
opposed to -- 1996! \cite{mandel}. 

Both Information Society and New Economy periods were not easy to 
understand for economists.  In a much quoted remark, in 1987
 Nobel Laureate Robert 
Solow said that ``we can see the computer age everywhere except in 
the productivity statistics'' (e.g.\ \cite{minehan}).  
A very gloomy view of the computing and telecoms sector was presented
in May 1998 by \cite{veneroso}: 
``Surveys now indicate that almost 50\% of all U.S.\  households own PCs. 
The PC is a sophisticated product. Educational levels, even literacy, are 
inadequate for a significant percentage of the U.S.  population. It is 
quite amazing that so high a percentage of all households own PCs. 
Clearly, market saturation, if it is not already here, cannot be far away.''
All one can say is, thank goodness new user interface technology saved
us all!  

An aspect of confusion for commentators on the technology swings has 
been the role of services versus manufactured goods.  
The problem with the following view of ICT application \cite{eib} is clear 
enough, namely that software and similar goods are in fact -- to an excellent 
approximation -- of zero cost from the second copy onwards: 
``intangible
goods, such as software and other digital information-goods, whose 
unit costs of production tend
to fall rapidly with growth in the volume of production.  
...
In this sense one may say that the information technology
revolution has been contributing towards maintaining the importance of 
the sector of the US
economy in which production is characterized by conventional, 
old-fashioned economies of scale.''  
So economy of scale is meant to explain what was happening in an ICT
upwelling.  I disagree: services, I believe, should not be distinguished
from other classes of goods.

Another perceived problem with understanding goods versus services 
is that true prices 
of services are difficult to fathom.   David \cite{eib} notes
``a substantial gap between average labour productivity
growth rates in the better-measured, commodity-producing 
sector on the one hand, and a collection
of `hard-to-measure' service industries, on the other.''

David \cite{eib} goes some way towards reconciling how tangible
goods can be influenced by intangible services.  He presents
``a view of the digital technology revolution as a source of 
efficiency improvements that gradually
have been increasing in magnitude and permeating the economy''.  
David sees computerization and telecoms as ``general purpose technologies''
or as a ``general purpose engine'', deployed in the framework of 
more established technologies.  He points to further progress to be 
expected, and as examples mentions (i) digitized and online intra-company
workflow, (ii) wearable and similar computing platforms; and (iii) 
tele-working.  However this perspective remains anchored in a view that 
efficiency of traditional market sectors is what is 
important, rather than something that is fundamentailly new. 

My view of this is quite different.  We are not witnessing  
just a ratcheting up 
of traditional efficiencies.  In fact what I cannot understand
is why a dividing line is drawn between goods and services.  When I 
hear of manufacturing being distinguished from services, as economic
categories, I am perplexed.  A manufactured good that does not 
perform a service when used or consumed -- that, to my mind, is a 
contradiction in terms.    If a service is purchased and 
consumed then surely that is immediately and directly a manufactured 
good.  

Indeed, further evidence of the tangibility of services was the
economic downturn triggered by the US subprime borrowing sector in 
the second half of 2007.  To illustrate how this can have 
implications for the ICT sector, consider 
 India's Tata Consultancy Services (TCS) \cite{tata}
with 65\% of its near US \$ 40 billion in revenues earned from the US. 
TCS's US earnings in turn accounted for a major share of its overall 
30\% of revenues from the banking and financial sector.  

Maybe changes of economic categories will come about.  
Industrial and sectoral categorization schemes must change over time.  
A standard is the NAICS, North American Industry Classification System
\cite{naics}.  The NAICS 2007 and the NAICS 2002 standards introduced a 
good number of changes, in particular in the ICT area.  Nonetheless 
there are high level categories for: Manufacturing, Information, 
Utilities, Professional (i.e.\ Services), and so on.  Links can be 
found at \cite{naics} also to the  North American Product Classification 
System (NAPCS) for services and separately for manufacturing.  
Industrial categorization schemes of these types have their use in 
particular areas.  An example of how a different scheme was developed 
is GICS, Global Industry Classification Standard.  It was developed 
by the financial sector (specifically Morgan Stanley Capital 
International and Standard and Poor's) to allow for a categorization 
that was better correlated with profitability and rate of growth.  

By distinguishing between a computer and telecom sector, on the one hand,
and others with which this sector interfaces, one really has to 
square lots of circles.  Consider the following ICT-related
categories, from \cite{jorgensen},  where for example
the software sector is cut off from its domain of application
(not a good idea from the software engineering viewpoint of user-centered
design). 
``IT-producing industries -- semiconductors, computers, communications 
equipment, and software ...
Although three-quarters of U.S. industries have contributed to the
 acceleration in economic growth, the four IT-producing industries 
are responsible for a quarter of the growth resurgence, but only 
3 percent of the Gross Domestic Product (GDP). IT-using industries 
account for another quarter of the growth resurgence and about the 
same proportion of the GDP, while non-IT industries with 70 percent 
of value-added are responsible for only half the resurgence. Obviously, 
the impact of the IT-producing industries is far out of proportion to 
their relatively small size.''  
Being stuck in an IT-producing versus IT-using view unfortunately
hinders greatly an understanding of the present or the near future.

I would propose that ``innovation'' has to be understood in 
conjunction with what is at issue,  
just as software is closely tied to its application.  
Unfortunately 
we must often 
discuss innovation in the abstract, and similarly the software sector in the 
abstract.  I have noted that 
instead it is the technology that has changed fundamentally.    
A useful supportive view is the following.  
While technological innovation, tax and deficit policy are interdependent, 
so that, for 
example, lower interest rates from increased savings can encourage 
innovation as can 
lower tax rates, nevertheless Mandel \cite{mandel} concludes that: 
``In the 1990s, at least, it seems 
that technology is more powerful than either taxes or deficits.''
It is my view that this is indeed the case, that any hard and fast
distinction between goods and services is unclear at best, and that 
software belongs to both camps.  

\subsection{The Financial Side of the New Economy}

The 1990s New Economy has been widely seen as an economic bubble
\cite{ghosh}.  
Two examples, among many, of how this worked in practice are as follows.
In what then as now is widely viewed as AOL's purchase of 
Time Warner in January 2000, both were 
roughly equally capitalized but there were 12,000-odd employees with the 
former and 67,000-odd with the latter.  
Another example of new buying old was in February 2000 when 
Vodafone (telecoms, mobile, UK-based)  bought out Mannesmann
(engineering, German-based).  By being massively valued, new wave ICT 
companies were able to buy out traditional, solid corporations
\cite{pennings}. 
In this section I probe the financial mechanism underlying this 
and its role 
in giving such strong trump cards to the new technologies.  

For Perez \cite{perez1,perez2}, booms such as the New Economy one are 
fueled by financial bubbles that are to be understood as 
 ``massive processes of credit creation'', ``massive episodes of 
credit creation''. 

Perez \cite{perez0} colorfully describes a financial and economic bubble 
as follows:  ``a whirlpool that sucks in huge amounts of the world's
wealth to reallocate it in more adventurous or reckless hands ... A 
part of this goes to new industries, another to expand the new 
infrastructure, another to modernize all the established industries,
but most of it is moved about in a frenzy of money-making money,
which creates asset inflation and provides a gambling atmosphere 
within an ever-expanding bubble''.  When new technologies that 
have instigated this have consolidated, a production phase sets 
in, and is viewed in far more favorable terms -- stable, 
equitable, just -- by Perez.  

Expressing the foregoing in another way, Perez points to the 
``techno-economic paradigm'' of development at issue here.  
Development of technology without finance is unthinkable. 
The causal connection between finance and technology is mutually
disruptive but simultaneously, at a deep level, constructive
and symbiotic.  Perez \cite{perez2} describes how: ``those radical 
innovative breaks also require bold and risk-loving bankers,
because the `serious' ones would share the same mental routines 
as the heads or managers of the established firms.  In fact, 
the historical recurrence of bursts of `wildcat or reckless' 
finance in the period of intense investment in technological 
revolutions suggests that these phenomena may be causally 
connected.''  

One other term used by Perez strikes a chord, that of ``clusters 
of radical innovation'',  
``Such interconnected innovations in products and processes, 
in equipment and organization, technical and managerial, form a 
coherent and mutually enhancing set of technologies and industries,
capable of carrying a wave of growth in the economy''.  
For we can see that in the earlier 1980s
Information Society boom that I discuss here, there was
the penetration of computerization into all aspects of business,
the rise of individual computing through the PC, Minitel as a 
precursor to society-wide networking, and various other facets.
In the 1990s New Economy boom there was a great surge in 
human-computer interface technology, mobile phone uptake soared, 
and industrial mergers of new and long-established partners took 
place, such as between 
AOL and Time Warner, or Vodafone and Mannesmann.

\section{The Changing Nature of the PhD}
\label{phd}

The PhD degree, including the title, the dissertation and the
evaluation framework 
as a work of research (the ``rite of passage'') 
came about in the German lands between the 1770s and the 1830s.  Clark 
\cite{clark} finds it surprising that it survived the disrepute
associated with all
academic qualifications in the turmoil of the late 18th century.  In the 
United States, the first PhD was awarded by Yale University in 1861.  In the 
UK, the University of London introduced the degree between 1857 and 1860.  
Cambridge University awarded the DPhil or PhD from 1882, and Oxford 
University only from 1917.  

\begin{figure}
\includegraphics[width=14cm]{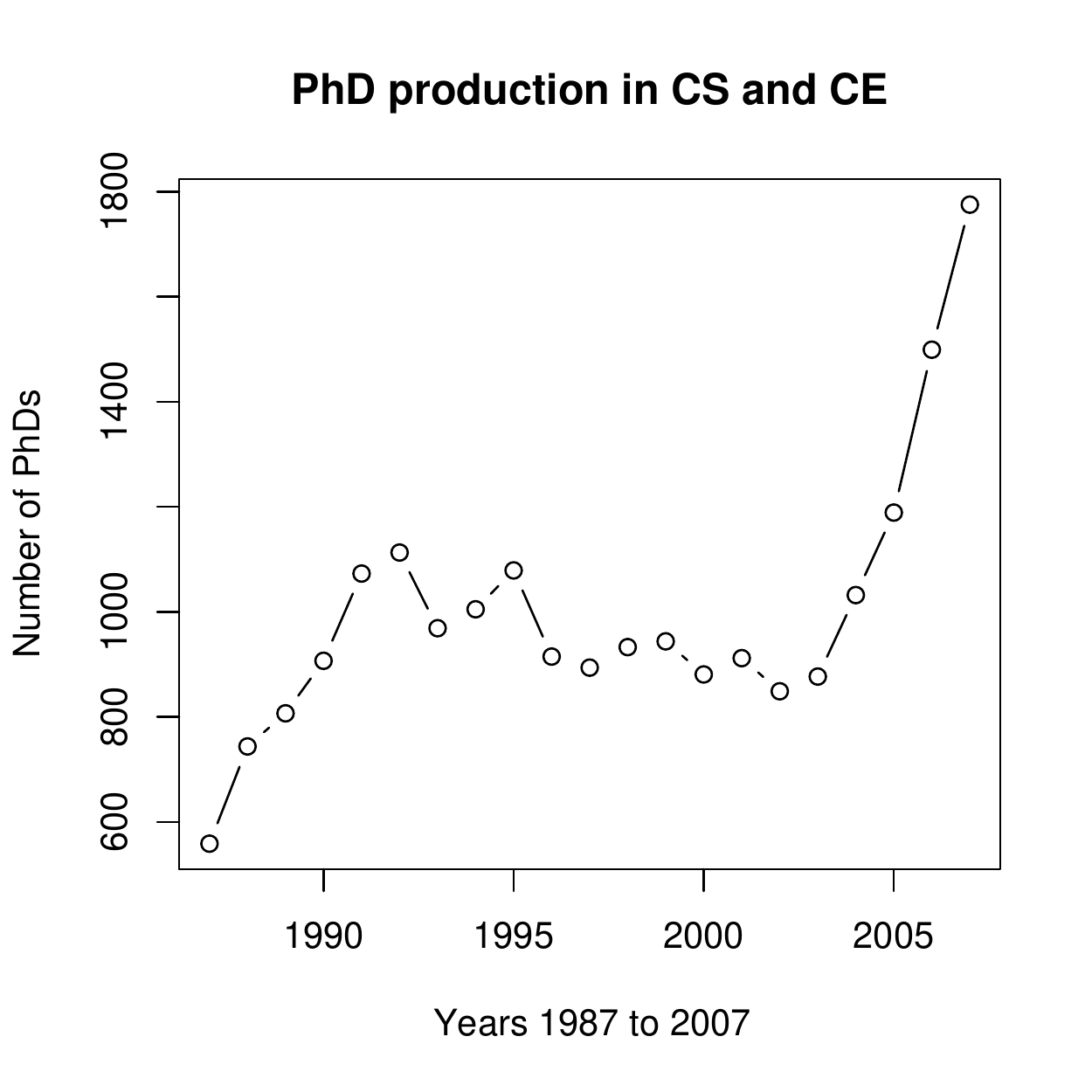}
\caption{PhD production in North America, in CS and CE.  From the 
CRA Taulbee Survey, May 2007.  While the growth may not continue, the 
increase in numbers of completing PhDs in CS and CE in the post Dot-Com 
or New Economy period is little short of stupendous.  For data see \cite{cra}.}
\label{fig5}
\end{figure}

A quite remarkable feature of the modern period, post Dot-Com 
or New Economy boom,
 is how spectacular the growth of PhD numbers has now become.  
Figure \ref{fig5} shows how PhDs dropped during the good years of the Dot-Com 
economy period.  But now the output trend in regard to PhDs 
is hugely different.  
In just four years, from 2003 to 2007, PhD output in CS and E 
in North America has 
doubled.   The Taulbee survey indicates that PhDs are expected to decline 
in the near future 
but by how much and whether then going into a further climb or a 
plateau are quite open issues.  

Internationally the evolution illustrated in Figure \ref{fig5} 
holds too.  For example, Ireland is pursuing a doubling of PhD output up 
to 2013 \cite{ssti}.  

Concomitant with numbers of PhDs, the very 
structure of the PhD is changing in many countries outside North 
America.  There is a strong movement away from the traditional
German ``master/apprentice'' model, towards instead a ``professional''
qualification.  This move is seen often as towards the US model.    
In Ireland there is a strong move to reform the PhD towards what is 
termed a 
``structured PhD''.  This involves a change from the apprenticeship model 
consisting 
of lone or small groups of students over three years in one university 
department to a 
new model incorporating elements of the apprenticeship model centered around 
groups of students possibly in multiple universities where generic and 
transferable 
skills (including entrepreneurial) can be embedded in education and 
training over four 
years.  

In Finland a Graduate School system was pursued on the cross-institutional 
level from 
1995.   Like the Irish case, the aim is for more systematic 
education and 
training that is more akin to that for a profession rather than as an 
apprenticeship.  An 
aim too is greater efficiency of advisor resource deployment and course 
provision, over a four year timeline \cite{researchfinland}.      
The Engineering Doctorate in the UK
is of similar duration, and professionally oriented \cite{epsrc1}. 
Doctoral Training Centres in the UK have similarities with 
the Graduate School concept
\cite{epsrc2}.   An
analogous situation holds for Graduate Schools in France, supporting a 
three year post-Master doctorate \cite{doctorat}. 
Unlike in these cases, Germany is retaining a traditional 
``master/apprentice'' model \cite{daad}.  

\begin{figure}
\includegraphics[width=14cm]{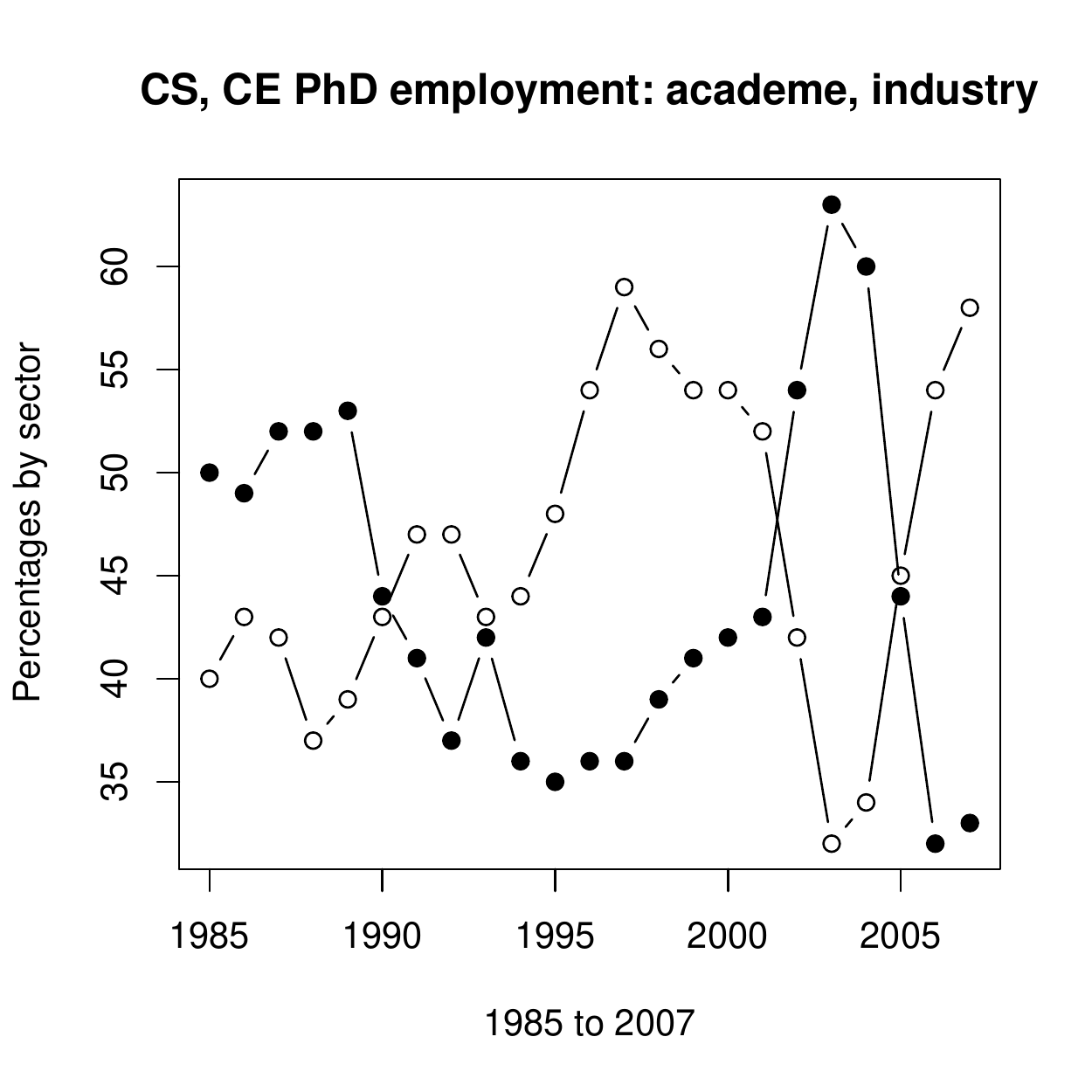}
\caption{Employment of PhDs produced in North America in CS and CE.  
From CRA Taulbee 
Survey, 2008 data. The curve with the filled dots represents 
academe, including in CS/CE awarding institute, other CS/CE institute, or 
non-CS/CE 
institute.  The curve with the open circles represents industry, including 
small 
percentages for government and self-employed.  Those who left North America 
are not included.  For data see \cite{cra}.}
\label{fig6}
\end{figure}

Numbers of PhDs are dramatically up, and as we have seen 
in many countries there is a 
major restructuring underway of the PhD work content and even timeline. 
In tandem with this, as Figure \ref{fig6} shows, in North America
the majority of PhDs now go directly into industry.  This trend goes 
hand in hand with the move from an apprenticeship for a career in 
academe to, instead, a professional qualification for a career in 
business or industry.

\section{Conclusion}

CS and E undergraduate recruitment and PhD production figures are 
all key data.  With various examples I have shown that they are 
also key to our understanding of a wide 
range of underlying social and technological trends.  


Using these key data to study underlying economical and technological 
changes ought not be left to others.  After all, we as 
Computer Scientists and Engineers have a better vantage point.  

The categories we use are supremely important.  The joint association of
computerization and telecoms in the term ICT is just one example.  
So too is the {\em multimedia information industry} \cite{pennings},
merging telecoms, information technology, entertainment, media and
consumer electronics.  
Official statistics lag very much behind this.  So facts and figures can 
mislead.  I have noted the confused overlapping terms ``manufacturing'' 
and ``services''.  An immediate conclusion is that policy makers can provide
leadership by using forward-reaching categorization and prioritization
 of research themes and directions.  
Steps in this direction are discussed further in \cite{boole}.

\end{document}